# Impurity-Induced Interference at a Topological Boundary in an Infinite SSH Heterojunction

Hao-Ru Wu, Yiing-Rei Chen and Hong-Yi Chen*
Department of Physics, National Taiwan Normal University, Taipei, 116, Taiwan

In this work, we investigate the coupling between a strong impurity and the topological boundary of an SSH heterojunction, composed of two SSH chains belonging to different topological classes. We show that impurity–boundary coupling gives rise to bonding and antibonding states within the SSH bulk gap. This coupling produces an interference effect in the local density of states, as the impurity approaches the boundary the LDOS evolves from a single sharp peak to a characteristic double-peak structure. Moreover, the interference strength can be quantified by the decay length of the bonding or antibonding wavefunction and by the energy splitting of the LDOS resonance peaks near the Fermi energy.

The discovery of topological insulators (TIs) has introduced the concept of symmetry-protected topological phases characterized by bulk topological invariants and the emergence of robust edge states [1–3]. In conventional approaches, impurities are often employed as probes to detect and characterize such edge states. In practical settings, however — most notably in scanning tunneling microscopy (STM) experiments — the measured signal cannot be used to unambiguously distinguish a topological edge state from an impurity-induced bound state without prior knowledge of the impurity strength [4,5]. This ambiguity motivates a detailed investigation of the interference between impurity states and topological edge modes.

Topological edge states are generally believed to be robust against disorder and local perturbations as long as the protecting bulk symmetries are preserved. Nevertheless, the deliberate introduction of defects or impurities into a topological system is of considerable practical interest, as such perturbations can locally disrupt the edge state [6,7]. The resulting in-gap resonances may, in turn, provide experimentally accessible signatures of nontrivial topology [8]. In particular, a single impurity has been proposed as a sensitive probe of topological order. In a trivial insulator, a weak impurity induces a bound state within the bulk gap that merges into the conduction band as the impurity strength increases. By contrast, in a topological insulator, a sufficiently strong impurity can effectively destroy the topologically protected edge state, locally creating a trivial region and leaving behind an impurity-bound state within the bulk gap. Such impurities may be treated either as hard-wall boundaries [9–11] or as discrete defects embedded in an otherwise clean system [6,12,13]; in both cases, an in-gap bound state emerges irrespective of the impurity strength.

Another closely related line of research concerns heterojunctions formed by two leads described by Su–Schrieffer–Heeger (SSH) models [14–17] or engineered graphene nanoribbon systems [4,5,18,19]. These junctions host localized boundary states whenever the difference in the topological invariants of the two leads (Fig. 1) is nonzero. This observation highlights that the bulk–boundary correspondence extends naturally to heterojunction geometries.

Motivated by these developments, we investigate a heterojunction composed of two semi-infinite SSH chains belonging to distinct topological phases, with a single impurity introduced at a distance $d$ from the junction interface. In the absence of the impurity, or when it is sufficiently far from the boundary, the local density of states (LDOS) at the interface exhibits a single peak within the bulk gap, corresponding to the topological boundary state. As the impurity approaches the boundary, hybridization between the impurity-induced bound state and the topological edge state leads to a characteristic splitting of this peak into two. This splitting provides a robust local indicator of the underlying topological state.

From an experimental perspective, STM measurements of impurity-induced bound-state energies offer the advantage of being insensitive to tunneling matrix elements and can yield quantitative information about impurity potentials [20–22]. Accordingly, we propose probing the LDOS via scanning tunneling spectroscopy (STS) at the boundary site. The observation of a double-peak structure in the LDOS serves as a clear experimental signature of the presence of a topological edge state.

* hongyi@ntnu.edu.tw

As illustrated in Fig. 1, our model is a heterojunction of two semi-infinite SSH chains. For both chains, we consider the same unit cell that contains a pair of $A$ and $B$ sublattices. The intra-cell hopping strength for the left(right) chain is $t_1(t_3)$, and the inter-cell hopping strength for both chains is $t_2$.

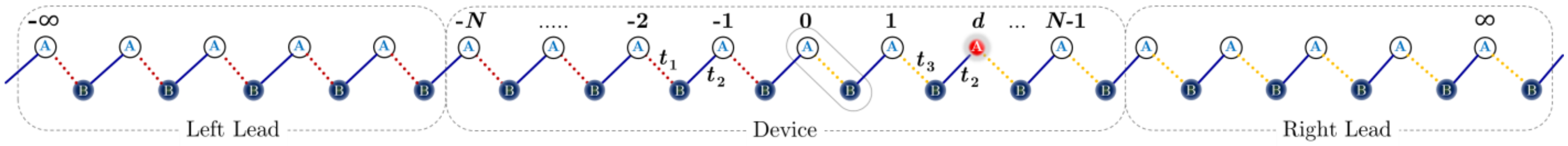

FIG. 1 (Color online) The two segments in the device, connected at the boundary and subjected to different intra-cell hoppings $t_1$ and $t_3$ (dotted lines), each reaches into the lead of its own semi-infinite extension. A single impurity of on-site potential $U$ is placed at the $A$-sublattice that is a distance $d$ away from the boundary at the $0^{\text{th}}$ unit cell (the 0-boundary).

The topological class for each semi-infinite chain can be obtained from its own infinite version, i.e., a 1D SSH system whose Hamiltonian takes the form

$$H^j(k) = \begin{pmatrix} 0 & h_k^j \\ h_k^{j^*} & 0 \end{pmatrix}, \qquad j \in L, R$$

where $h_k^L = t_1 + t_2 e^{ik}$ (or $h_k^R = t_3 + t_2 e^{ik}$) is used for the left (or right) chain and indexed by the 1D crystal momentum $k$. The topological properties of such infinite 1D systems can be characterized by the winding number, which is given by [23]

$$\gamma_j = \frac{1}{2\pi i} \oint \partial_k \log\left[h_k^j\right] dk,$$

where the integral can be obtained by counting the number of times $h_k^j$ goes around the origin in the complex plane. The winding number of such an infinite SSH system can be determined by the ratio of the intra-cell and inter-cell hopping strengths as [14, 24],

$$\gamma_L = \begin{cases} 0, & t_1/t_2 > 1 \\ 1, & t_1/t_2 < 1 \end{cases}$$

$$\gamma_R = \begin{cases} 0, & t_3/t_2 > 1 \\ 1, & t_3/t_2 < 1 \end{cases}$$

In particular, it is not the exact value of the winding number that matters anything to the localized boundary states. The difference of the winding number $|\gamma_L - \gamma_R|$ indicates the emergence of localized states at boundaries due to the bulk-boundary correspondence [25].

We then adopt the Green's function (GF) method, and divide our model system into 3 parts, namely the left lead, the right lead, and the device. The left (right) lead is a semi-infinite 1D SSH chain coming all the way from $-\infty(\infty)$ to meet the device (Fig. 1). The device contains 2 connected segments, where the left (right) segment is just a continuation of the left (right) lead. We spare plenty of cells in both segments, so the device serves as a window that facilitates LDOS observation for the boundary zone. The surface Green's functions [26] of both leads are calculated iteratively by

$$\boldsymbol{g}_{sL}(E) = \left[\boldsymbol{\alpha}_L - \boldsymbol{\beta}_L \boldsymbol{g}_{sL}(E) \boldsymbol{\beta}_L^\dagger\right]^{-1},$$

$$\boldsymbol{g}_{sR}(E) = \left[\boldsymbol{\alpha}_R - \boldsymbol{\beta}_R \boldsymbol{g}_{sR}(E) \boldsymbol{\beta}_R^\dagger\right]^{-1},$$

where $\boldsymbol{\alpha} = (E + i\eta - \mathbf{h})$ and $\mathbf{h}$ is the Hamiltonian matrix of each unit cell, and $\boldsymbol{\beta}$ is the coupling matrix between neighboring cells as

$$\mathbf{h}_L = \begin{pmatrix} 0 & t_1 \\ t_1 & 0 \end{pmatrix}, \qquad \boldsymbol{\beta}_L = \begin{pmatrix} 0 & t_2 \\ 0 & 0 \end{pmatrix},$$

$$\mathbf{h}_R = \begin{pmatrix} 0 & t_3 \\ t_3 & 0 \end{pmatrix}, \qquad \boldsymbol{\beta}_R = \begin{pmatrix} 0 & 0 \\ t_2 & 0 \end{pmatrix}.$$

With the device's Hamiltonian matrix being $\mathbf{H}_D$, the effective GF, in the device, is therefore

$$\mathbf{G}_{\text{eff}}(E) = \frac{1}{E + i\eta - \mathbf{H}_D - \boldsymbol{\beta}_L \boldsymbol{g}_{sL}(E) \boldsymbol{\beta}_L^\dagger - \boldsymbol{\beta}_R \boldsymbol{g}_{sR}(E) \boldsymbol{\beta}_R^\dagger}$$

Certainly, the impurity is embedded in the device, in the vicinity of the boundary. The device's Hamiltonian $H_{\text{D}}$ includes the two different SSH segments, as well as the impurity:

$$H_D = H_0 + H_{\text{imp}},$$

where

$$H_0 = \left[\sum_{i=-1}^{-N} t_1 c_{i,A}^\dagger c_{i,B} + \sum_{i=0}^{N-1} t_3 c_{i,A}^\dagger c_{i,B} + \sum_{i=-(N-1)}^{N-1} t_2 c_{i,A}^\dagger c_{i-1,B}\right] + \text{h.c.},$$

$$H_{\text{imp}} = U c_{d,A}^\dagger c_{d,A},$$

$c_{i,\alpha}^\dagger$ ($c_{i,\alpha}$) is the creation (annihilation) operator of sub-lattice $\alpha$ ($\alpha = A$, or $B$) in the $i^{th}$ unit cell, $U$ is the on-site repulsive potential of the impurity, and $d$ is the impurity's distance from the boundary. Note that in this study, we reserve $N = 40$ for each segment. While we have chosen $t_2 \equiv 1$ to scale all other energies, a strong impurity[1] $U = 100$ is used throughout this paper. The imaginary part of the effective GF, namely $\mathbf{D}(E) \equiv -\frac{1}{\pi}\mathrm{Im}\,\mathbf{G}_{\mathrm{eff}}(E)$, is the matrix where the LDOS of the $i^{th}$ unit cell can be directly read off from the diagonal (onsite) matrix elements:

$$\rho_{i,\alpha}(E) \equiv D_{(i,\alpha)(i,\alpha)}(E)$$

For the impurity-free limit, i.e., $U = 0$, it is known that a boundary state can occur in such a heterojunction, depending on the values of $t_1$ and $t_3$. As $t_1 > t_2$ and $t_3 < t_2$, the boundary state appears at the site $(0, A)$, while as $t_1 < t_2$ and $t_3 > t_2$, the boundary state appears at the site $(-1, B)$. By keeping track of $\rho_{0,A}$ and $\rho_{-1,B}$, right on the Fermi energy ($E_F = 0$), we show the trend of boundary state's occurrence in the parameter space of $t_1$ and $t_3$ (Fig. 2(a)). In the upper-right and lower-left quadrants, there is no winding number difference $|\gamma_L - \gamma_R| = 0$, an energy gap of $2\min(|t_1 - t_2|, |t_3 - t_2|)$ sits right on $E_F$ and $\rho_{0,A}(E_F) = \rho_{-1,B}(E_F) = 0$ indicates the absence of the topological boundary state. For example, the parameter set $(t_1, t_3) = (0.4, 0.5)$, corresponding to point $p_1$ in the lower-left quadrant, yields a boundary LDOS without signatures of a topological boundary state. In contrast, the parameter set $(t_1, t_3) = (1.5, 0.7)$, corresponding to point $p_2$ in the upper-left quadrant where $|\gamma_L - \gamma_R| = 1$, exhibits a pronounced peak in $\rho_{0,A}(E_F)$ (see Fig. 2(c)), providing clear evidence of a topological boundary state.

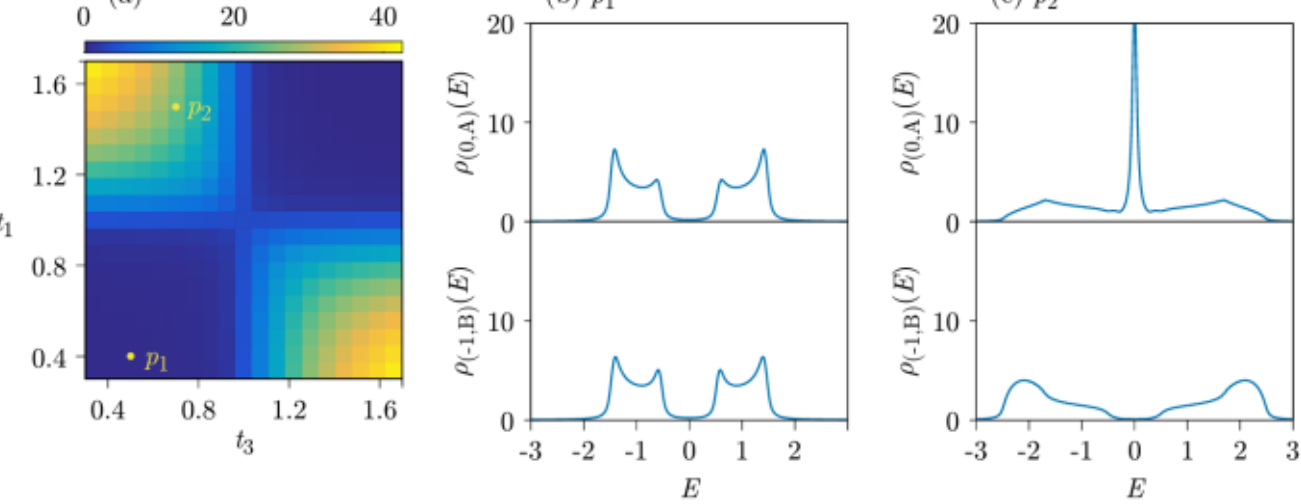

FIG. 2. (Color online) (a) The LDOS at the boundary, $\rho_{0,A}(E_F) + \rho_{-1,B}(E_F)$ in the parameter space of hopping parameters $t_1$ and $t_3$. The upper-right and lower-left quadrants are topologically trivial, while the upper-left and lower-right quadrants are topologically non-trivial. $p_1$ and $p_2$ correspond to hopping parameter sets $(t_1, t_3) = (0.4, 0.5)$ and $(1.5, 0.7)$, respectively. (b) and (c) are two typical cases from each regime, whose $\rho_{0,A}(E)$ and $\rho_{-1,B}(E)$ plots are shown separately.

Adding one impurity to this heterojunction system leads to the following interesting observations that bring out the properties embedded in the topology framework. Consider a single impurity planted on an A sublattice on the right-hand-side segment, avoiding the boundary. In the topological regime ($|\gamma_L - \gamma_R| = 1$), the impurity couples to the boundary state, and the coupling depends on the SSH model profile. However, such coupling does not happen in the non-topological regime ($|\gamma_L - \gamma_R| = 0$). Fig. 3 shows the effect due to the impurity in two categories of heterojunctions. In a topological junction, the well-defined $\rho(E)$ peak at $E_F$ evolves as the impurity is placed closer to the boundary, and becomes a pair of well-separated peaks inside the bulk gap. As shown in Fig.3(a)-(c). In a topologically trivial junction, the energy gap in the LDOS remains untouched as the impurity is placed to approach the boundary, as shown in Fig. 3(d)-(f). The impurity, kept away from the boundary by $d \neq 0$, does not destroy the bulk-boundary correspondence, and indeed introduces peculiar interference effect.

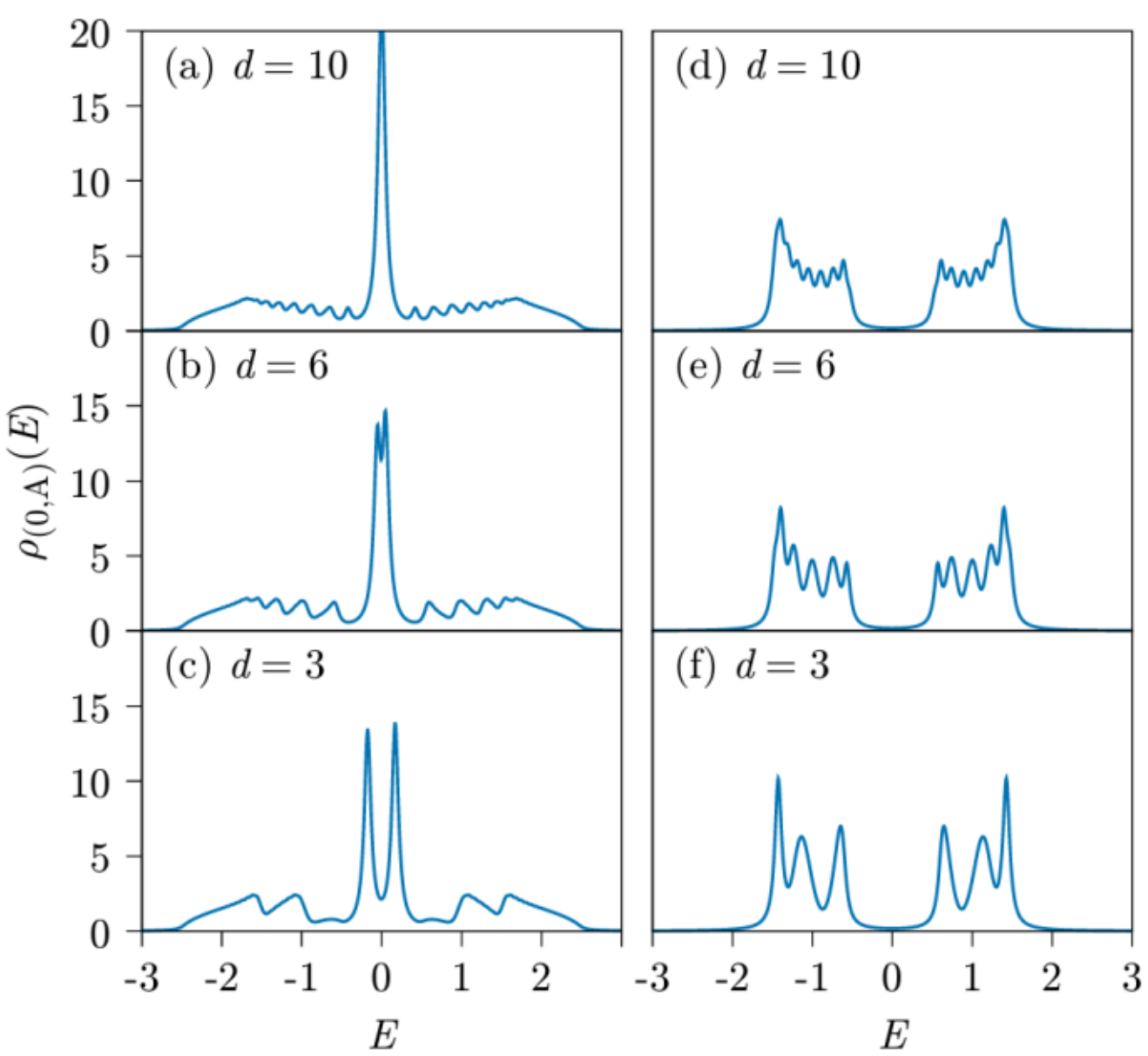

FIG. 3. Left: The $\rho(E)$ plots for impurity planted away from the boundary at distance (a) $d = 10$, (b) $d = 6$ and (c) $d = 3$, in an infinite heterojunction of $t_1 = 1.5$, $t_3 = 0.7$ and $|\gamma_L - \gamma_R| = 1$. Right: Same series of observations in an infinite heterojunction of $t_1 = 0.4$, $t_3 = 0.5$ and $|\gamma_L - \gamma_R| = 0$.

[1] For this article's major theme of interference phenomena that will soon be revealed in the discussions that follow, despite that the choice of repulsive ($U > 0$) or attractive ($U < 0$) impurity does leads to opposite energy shifting that accompanies energy splitting, and certainly brings an overall phase and/or differences in wave-function composition profile, truly it affects neither the interference phenomena, nor our conclusions deduced thereon.

To comprehensively account for the splitting of the zero-energy peak in $\rho(E_F)$ and hence the impurity-boundary coupling, we study the exact diagonalization on a finite SSH heterostructure with $N = 200$ for each segment, which not only gives the similar boundary LDOS results as the infinite heterojunction, but also provides specific electronic wavefunctions in real space of the lattice sites. Two finite SSH heterostructures are promising choices: a linear structure and a ring structure. The linear structure is simply the device portion, disconnected from the leads and left with two free ends. The ring structure can be formed by connecting the free ends of the two SSH segments in the device, originally connected to the semi-infinite leads, as $t_2 \hat{c}^{\dagger}_{-N,A} \hat{c}_{N-1,B} + \mathrm{h.c.}$. In this way, we create one other boundary (N-boundary) that sits opposite to the original one (0-boundary), and the device's loose ends connect to form a ring.

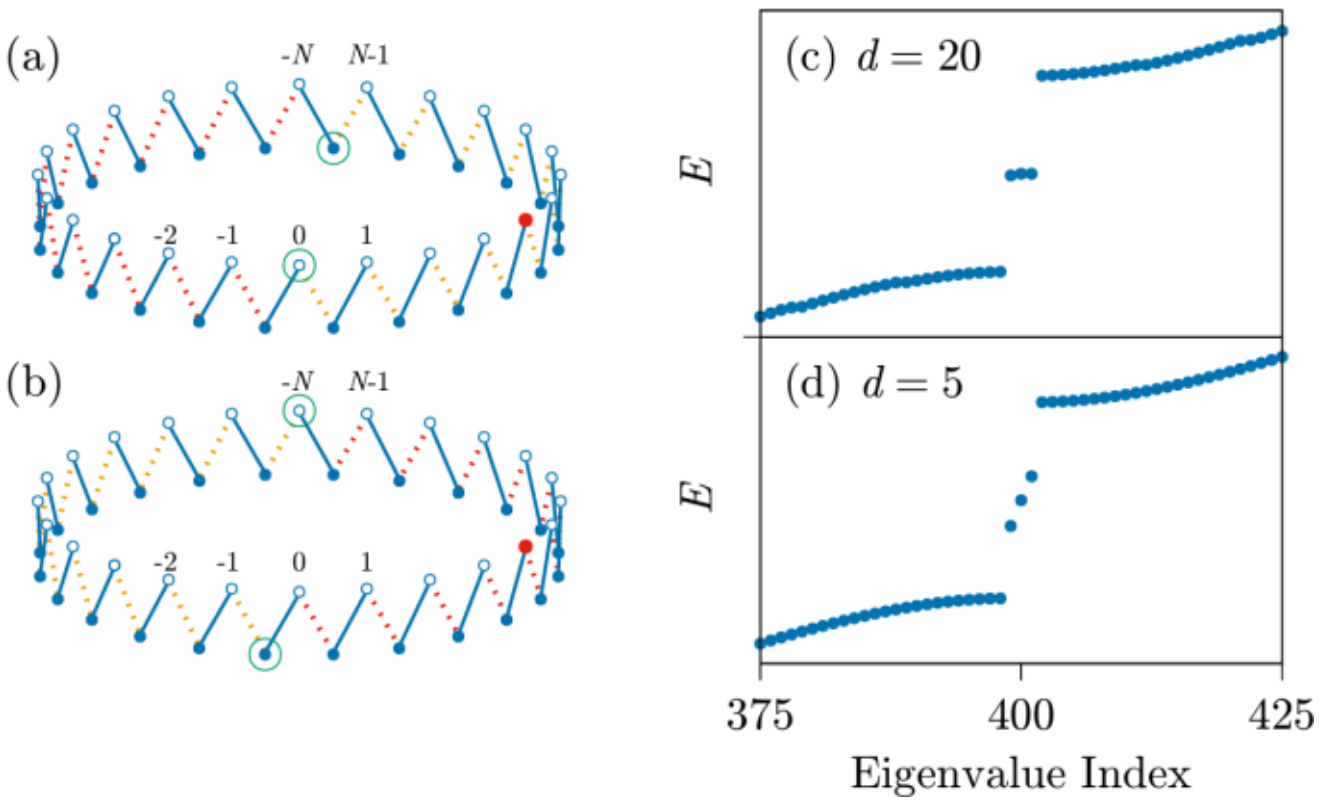

FIG. 4. (Color online) Left: Ball-and-stick illustrations for (a) $t_1 > t_2, t_3 < t_2$, the 0-boundary sits on an A-cusp and the N-boundary on a B-cusp; and (b) $t_1 < t_2, t_3 > t_2$, the 0-boundary sits on a B-cusp and the N-boundary on an A-cusp. Right: The energy eigenvalues of the topological ring for (c) $d = 20$ and (d) $d = 5$ with hopping parameters $t_1 = 1.5$ and $t_3 = 0.7$. The bulk bandgap is $0.6$, $N = 200$.

In all cases we consider for the infinite heterojunction and study with the GF method, the LDOS features we identify therein are present in both finite structures, linear and ring-shaped. In the following discussions, we shall exclusively use the SSH heterojunction ring to present our results. The peculiar details, especially the in-gap states due to the linear structure's unique open ends, are addressed in the Appendix, in comparison with the ring.

Using the ring as a working system, however, one notices at once that it has two boundaries. In the topologically non-trivial regime and impurity-free limit, the ring gives two boundary states, instead of one. Moreover, due to the ring structure, these two boundary states, sitting opposite to each other, always occupy different sublattices, where the "cusps" occur in the SSH model. When the 0-boundary sits on the A-cusp, the N-boundary surely sits on the B-cusp corresponding to the upper-left topological quadrant, and vice versa (Fig. 4(a) and 4(b)).

Fig. 4(c) presents energy eigenvalues of one impurity planted in the topological SSH heterojunction ring. In the case of $d = 20$ where the two boundaries are equidistant from the impurity, there are three in-gap states: $E = 0^+$, 0, and $0^-$. The state of $E = 0$ is simply the N-boundary state, which locates exclusively on the B-sublattice of the N-boundary (Fig. 4(a)). The state of $E = 0^+$ or $E = 0^-$ indicates the interference between 0-boundary and impurity. The wavefunction depicted in Fig. 5(a) and 5(c) shows that state $E = 0^+$ and state $E = 0^-$ are a pair resulting from interference, in fact the anti-bonding state ($|\psi_{\text{antibonding}}\rangle = |\psi_{\text{0-boundary}}\rangle - |\psi_{\text{imp}}\rangle$) and bonding state ( $|\psi_{\text{bonding}}\rangle = |\psi_{\text{0-boundary}}\rangle + |\psi_{\text{imp}}\rangle$ ) between the 0-boundary and the impurity. The wavefunction of the state $E = 0^+$ distributes mainly on the 0-boundary, i.e., $|\psi_{\text{antibonding}}\rangle \approx |\psi_{\text{0-boundary}}\rangle$, while the wavefunction of the state $E = 0^-$ distributes primarily on the impurity spots, i.e., $|\psi_{\text{bonding}}\rangle \approx |\psi_{\text{imp}}\rangle$. Thus, the interference is so weak that the LDOS $\rho_{0,A}(E)$ exhibits single peak.

In the case of $d = 5$, the three in-gap eigenvalues are $E = 0.07$, 0, and $-0.07$ (Fig. 4(d)). The state of $E = 0$, i.e., the N-boundary is curiously left out of any interference. As $d$ is decreased, the interference is strengthened, energy levels $E = 0^-$ and $E = 0^+$ (bonding state and anti-bonding state) split further, their wavefunctions distribute more competitively between the 0-boundary and the impurity (Fig. 5(d) and 5(f)). Thus, the LDOS $\rho_{0,A}(E)$ clearly exhibits double-peak structure.

It is worth noting that, within the upper-left topological quadrant of Fig. 2(a), an A-sublattice impurity couples to the 0-boundary located at the A-cusp, whereas a B-sublattice impurity instead couples to the boundary at the B-cusp, i.e., the N-boundary.

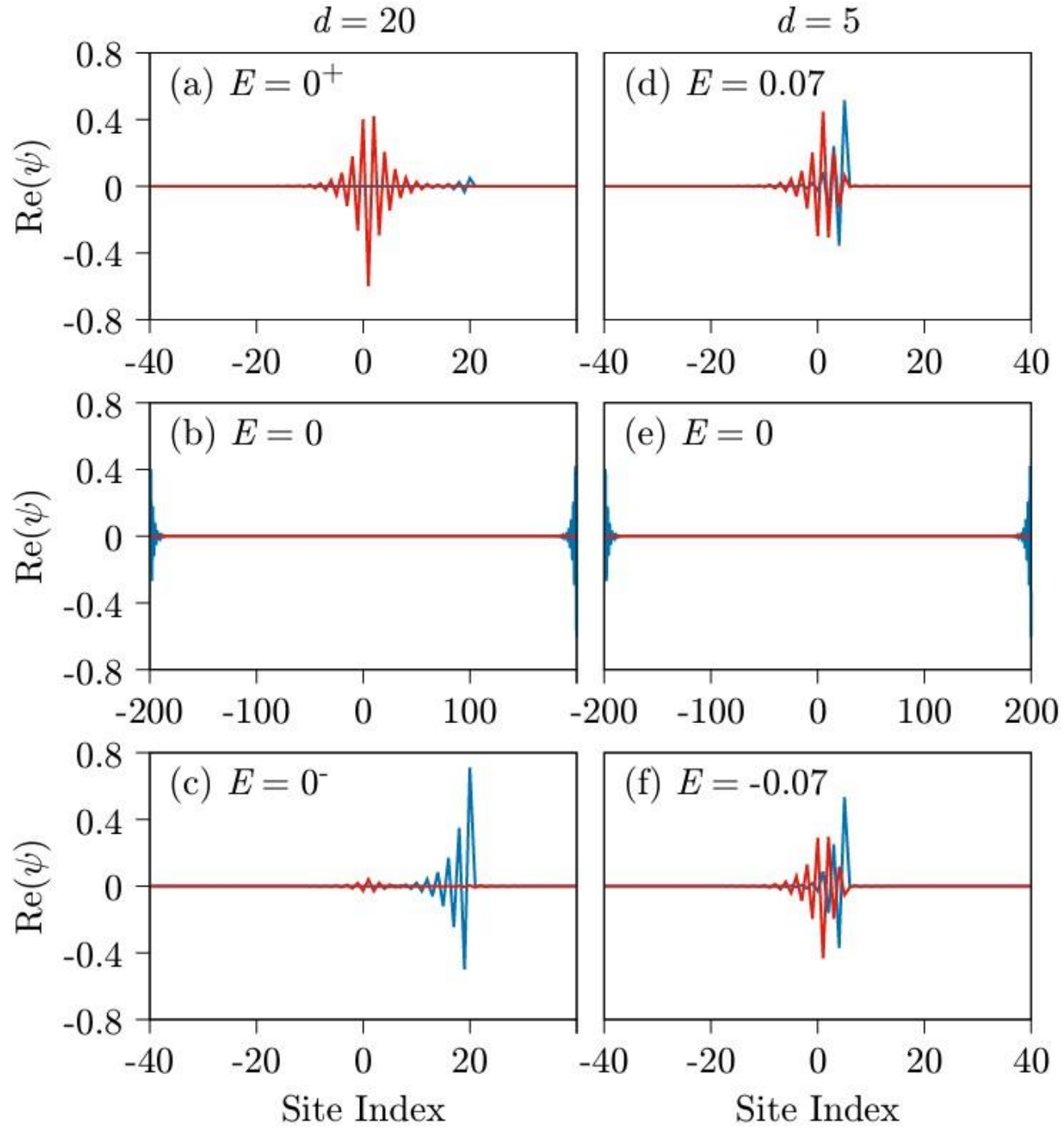

FIG. 5. (Color online) Real part of the in-gap states' wavefunctions, shown in real space of the lattice sites. (a)-(c) illustrate the case with the impurity at $d = 20$, and (d)-(f) are for the case with the impurity at $d = 5$ with hopping parameters $t_1 = 1.5$ and $t_3 = 0.7$. For each $d$, the 3 states shown here correspond to the in-gap dots in Fig. 4. The red(blue) line represents the projection on the $A(B)$ sublattices.

Since the inter-cell and intra-cell hoppings determine the bulk gap of a SSH chain, as well as the characteristic length that governs the profile of a local wavefunction [23] in the chain, the impurity-boundary coupling should certainly reveal the influence from the configuration of these hopping parameters. Take the coupling between an A-sublattice impurity and the 0-boundary for example. The decay of the A-sublattice portion of their bonding state is featured by the characteristic length $\xi$ in the exponential decay trend of $e^{-|x_i - d|/\xi}$, where $x_i$ is the position of atomic sites in the region between 0-boundary and the impurity. Going over the upper-left ($t_1 > 1, t_3 < 1$) topological regime, we find the relation between $\xi$ and the hopping parameter configuration: $\xi = 1/|\ln(t_2/t_3)|$ (Fig. 6(a)). It is worth to note that in the same topological regime, as $d < 0$ the decay length becomes $\xi = 1/|\ln(t_2/t_1)|$.

The energy splitting between the bonding and anti-bonding states is another signature of the interference. Fig. 6(b) shows $\Delta E = E^+ - E^-$ vs. impurity-boundary distance $d$. As shown by the line of trend in the figure, the points go like $\Delta E \sim \exp[-(d-1)/\xi]$ at small $d$ and saturates to a finite number at large $d$. As the impurity strength $U$ approaches infinity, the trend becomes linear with increasing $d$.

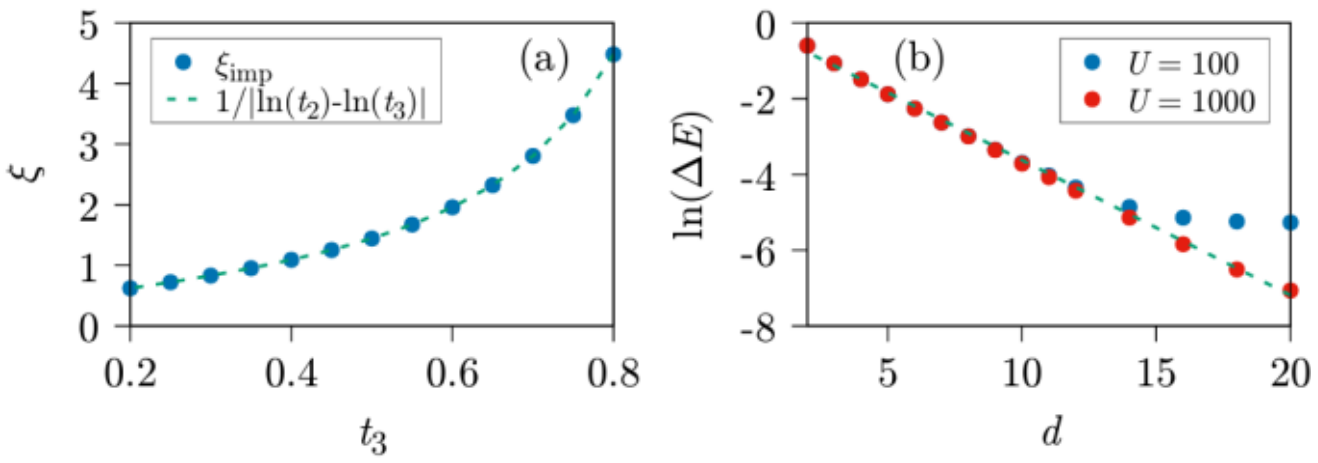

FIG. 6. (Color online) (a) For different $t_3$, the decay length $\xi$ of the bonding state between the 0-boundary and the impurity planted on the $A$-sublattice at $d = 20$. Blue dots: the numerical results. Dashed line: $\xi = 1/|\ln t_2 - \ln t_3|$. (b) The energy splitting $\Delta E = 0^+ - 0^-$ versus distance $d$ for $t_1 = 1.5$ and $t_3 = 0.7$. The relation between $\Delta E$ and $d$ goes like $\Delta E \sim \exp[-(d-1)/\xi]$, as marked by the dashed line.

To conclude, we have studied an infinite SSH heterojunction to explore impurity–boundary interference effects. Owing to bulk–boundary correspondence, a clean heterojunction composed of two topologically inequivalent materials exhibits a well-defined LDOS peak at the boundary. When an impurity couples to the boundary, bonding and antibonding states emerge, and their interference leads to a splitting of the corresponding energy levels. As a result, the single LDOS peak evolves into a distinct double-peak structure. The interference strength can be quantified by both the decay length of the wavefunctions and the energy splitting. Specifically, the decay length follows a logarithmic dependence on the hopping parameters, while the energy splitting of the double-peak LDOS varies linearly with the impurity–boundary distance $d$ in the limit $U \to \infty$. Our findings provide a clear and experimentally distinguishable signature for identifying a topological boundary.

## ACKNOWLEDGEMENTS

H.Y.C. was supported by Intelligent Computing for Sustainable Development Research Center at NTNU. Y.R.C thanks the support from National Science and Technology Council, grant No. NSTC 114-2635-M-003-001-MY3.

## APPENDIX

This section includes the comparison between the finite ring heterojunction (ring for short) and the finite linear heterojunction (line for short), for cases of three winding number configurations (i) $\gamma_L = \gamma_R = 0$, (ii) $\gamma_L =$ $\gamma_R = 1$ and (iii) $\gamma_L = 1$, $\gamma_R = 0$ (or $\gamma_L = 0$, $\gamma_R = 1$). It is meant to give further insight into the difference between the two structures, and account for our choice of the ring in the main text.

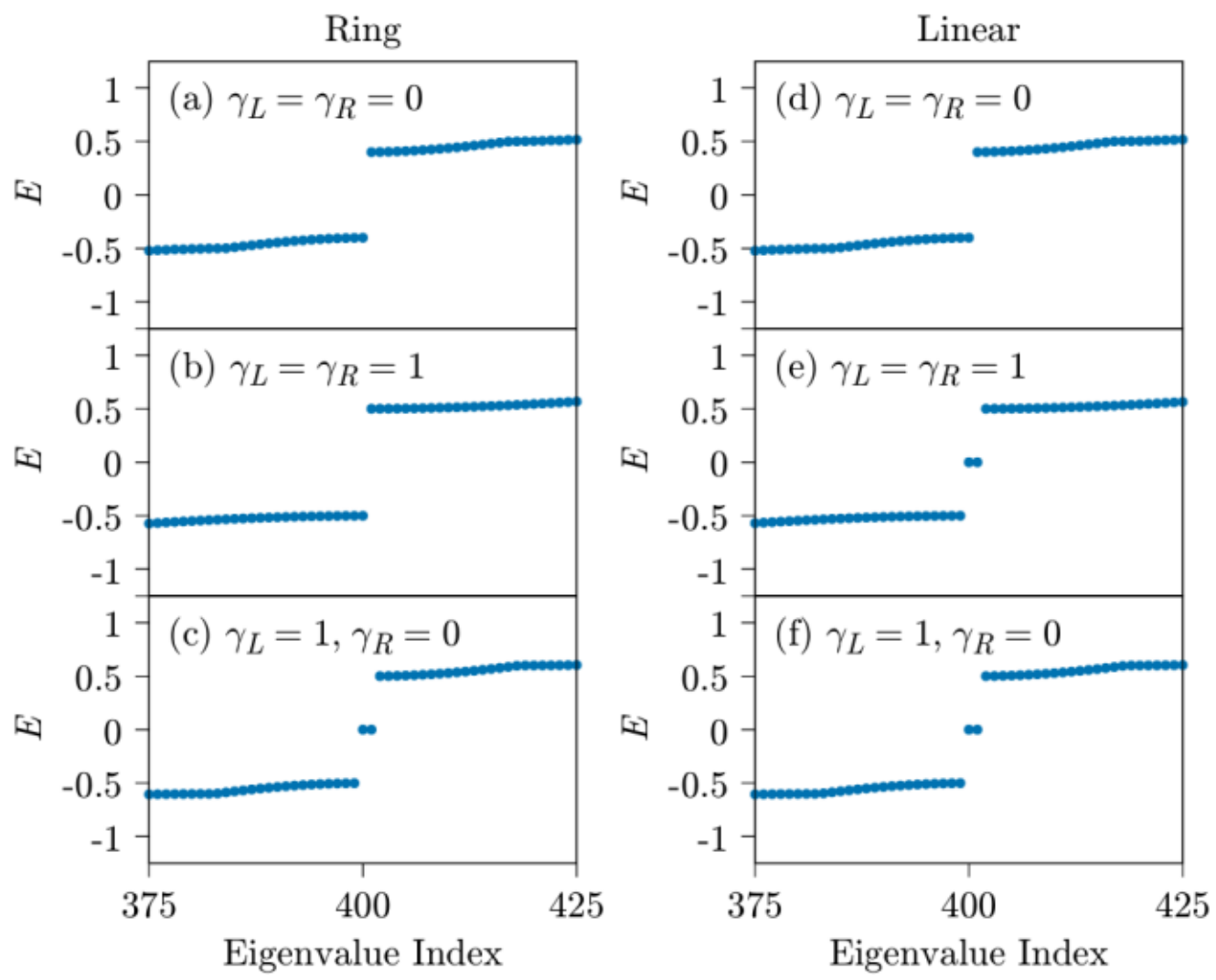

FIG. a. The energy spectra of the ring and the line, for the three winding number configurations.

As shown in Fig. a, the energy spectrum of the ring and that of the line exhibit slight but crucial differences. Note that case (i) and case (ii) both belong to the non-topological regime (in the upper-right and lower-left quadrants of Fig.2). While the ring's energy spectra for both cases are fully gapped, the line does have two in-gap states in case (ii). Unique to the finite line structure, these two energy levels correspond to edge states localized at the two open ends, rather than boundary states that should sit right on the junction.

In case (iii), which belongs to the topological regime, both structures host two in-gap states. For the ring, these two states are boundary states localized at the two heterojunction interfaces that are seated on sublattice A and sublattice B, respectively. For the line, one of the in-gap states corresponds to a heterojunction boundary, while the other one sits on an open end of the SSH segment with winding number $\gamma = 1$.

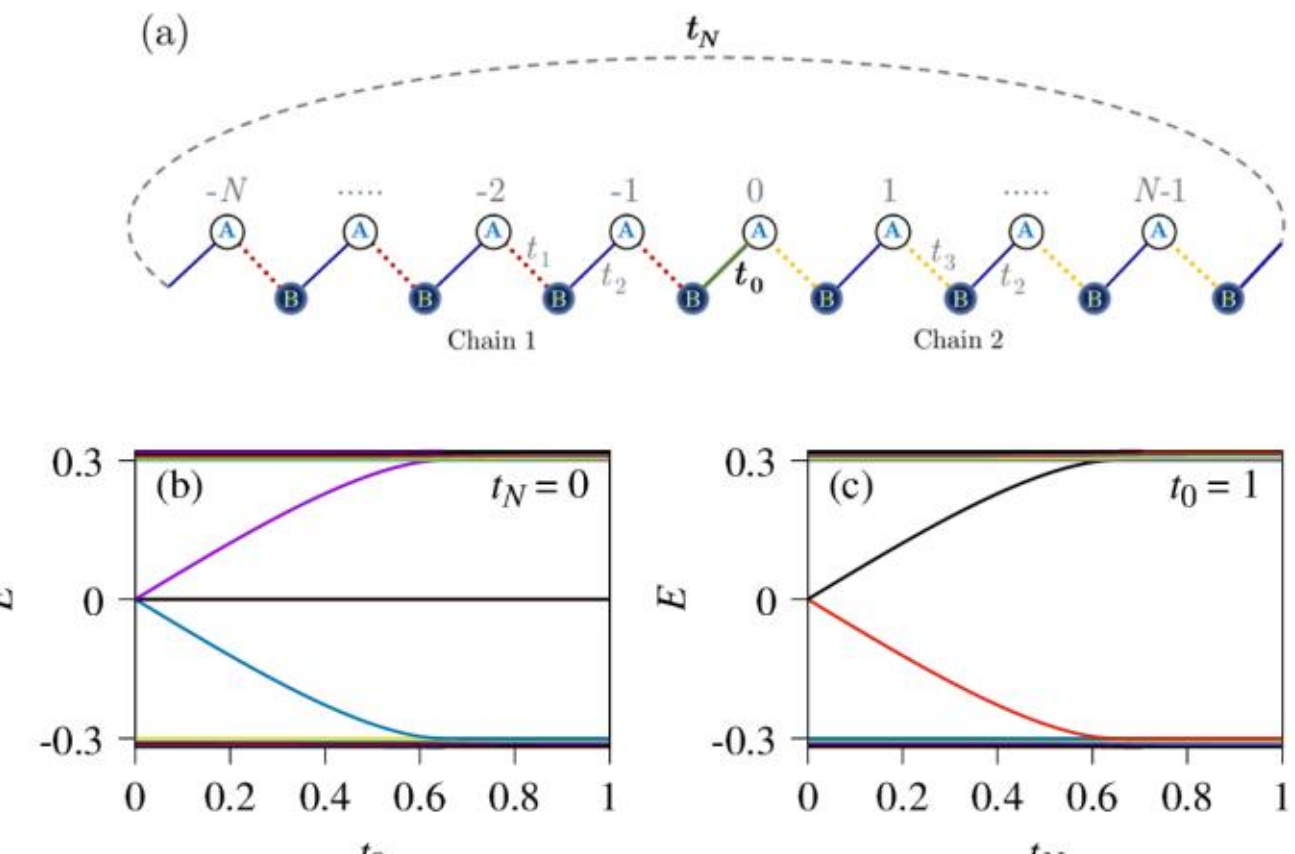

FIG. b. Evolution of the open ends' energy spectrum, as hopping strength parameters $t_0$ and $t_N$ are tuned up in orders.

It is worth pointing out that the difference between the ring and the line energy spectra for case (ii), i.e., $\gamma_L = \gamma_R = 1$, originates from the hybridization of edge states. To illustrate our point, we start from two decoupled SSH chains, each with winding number $\gamma = 1$, as shown in figure A2. In this limit, there are altogether four in-gap states, localized at the four open ends of the two chains. We then introduce a tunable hopping strength $t_0$ that gradually connects the two chains and form one boundary (so the line structure is formed), following which we bring in another tunable hopping strength $t_N$ that gradually assembles yet another boundary (so the ring is formed).

As we tune up $t_0$, the two edge states (of the gradually connected two end points) start to hybridize and their energies shift into the bulk bands, while the other two edge states remain in the gap. When we reach $t_0 = 1$, the line structure is completed. As we go forward to tune up $t_N$, the remaining two edge states also start to hybridize and their energies move into the bulk region as well. When we reach $t_N = 1$, the ring structure is completed, and no in-gap states remain.